\documentstyle[12pt]{article}
\catcode`@=11
\def\citen#1{\if@filesw \immediate\write \@auxout {\string\citation{#1}}\fi%
\@tempcntb\m@ne \let\@h@ld\relax \def\@citea{}%
\@for \@citeb:=#1\do {\@ifundefined {b@\@citeb}%
    {\@h@ld\@citea\@tempcntb\m@ne{\bf ?}%
    \@warning {Citation `\@citeb ' on page \thepage \space undefined}}%
    {\@tempcnta\@tempcntb \advance\@tempcnta\@ne
    \setbox\z@\hbox\bgroup\ifcat0\csname b@\@citeb \endcsname \relax
    \egroup \@tempcntb\number\csname b@\@citeb \endcsname \relax
    \else \egroup \@tempcntb\m@ne \fi \ifnum\@tempcnta=\@tempcntb
    \ifx\@h@ld\relax \edef \@h@ld{\@citea\csname b@\@citeb\endcsname}%
    \else \edef\@h@ld{\hbox{--}\penalty\@highpenalty
    \csname b@\@citeb\endcsname}\fi
    \else \@h@ld\@citea\csname b@\@citeb \endcsname \let\@h@ld\relax \fi}%
\def\@citea{,\penalty\@highpenalty\hskip.13em plus.13em minus.13em}}\@h@ld}
\def\@citex[#1]#2{\@cite{\citen{#2}}{#1}}%
\def\@cite#1#2{\leavevmode\unskip\ifnum\lastpenalty=\z@\penalty\@highpenalty\fi%
   $^{\scriptscriptstyle \multiply\@highpenalty 3 \mbox{\rm\scriptsize#1%
  \if@tempswa,\penalty\@highpenalty\ #2\fi}}$}   %

\def\dcite{\@ifnextchar [{\@tempswatrue\@dcitex}{\@tempswafalse\@dcitex[]}}

\def\@dcitex[#1]#2{\if@filesw\immediate\write\@auxout{\string\citation{#2}}\fi
  \def\@dcitea{}\@dcite{\@for\@dciteb:=#2\do
    {\@dcitea\def\@dcitea{,}\@ifundefined
       {b@\@dciteb}{{\bf ?}\@warning
       {d(line)cite  `\@dciteb' on page \thepage \space undefined}}%
\hbox{\csname b@\@dciteb\endcsname}}}{#1}}

\def\@dcite#1#2{$\mbox{\rm#1\if@tempswa , #2\fi}$}
\catcode`@=12

\def\sun{{\rm SU}(N)}
\def\spn{{\rm Sp}(N)}
\def\son{{\rm SO}(N)}
\def\cS{{\cal S}}
\def\tcS{{\tilde{\cal S}}}
\def\cK{{\cal K}}
\def\cZ{{\cal Z}}

\def\cW{{\cal W}}
\def\cY{{\cal Y}}

\def\Eu{{\cal E} }
\def\exp{{\rm exp}}
\def\bU{ { U_1,\ldots, U_b } }
\def\RS{ {  \dot{R} S   } }
\def\dP{ \dot{P}}
\def\dR{ \dot{R}}
\def\cV{{\cal V}}

\def\lam{\lambda}
\def\kap{\kappa}
\def\smb{{\footnotesize \#}}
\def\hchi{{\raise 0.8mm \hbox{$\chi$}}}
\def\eq{\begin{equation}}
\def\en{\end{equation}}
\def\sst{\scriptscriptstyle}
\def\sct{\scriptstyle}
\def\ds{\displaystyle}
\def\hep#1#2#3#4{hepth$\#$}
\def\multip{ {N_{\dR_1S_1, \dR_2S_2}}^{\dR_3S_3} }
\begin{document}
\setlength{\baselineskip}{24pt}
\thispagestyle{empty}

\hfill              \begin{tabular}{l} {\bf \hep-th/9411143} \\
                                        {\sf BRX-TH--359} \\
                                        {\sf BOW-PH-103}
                                     \end{tabular}

\vspace{1.7cm}

\begin{center}
\begin{tabular}{c}
{\LARGE The String Calculation of QCD Wilson Loops} \\[0.3cm]
{\LARGE    on Arbitrary Surfaces}
\end{tabular}
\vspace{1.5cm}
\end{center}
\setcounter{footnote}{2}

\begin{center}
{\Large Stephen G. Naculich$^\dagger$ and
        Harold A. Riggs\footnote{Supported in part by  DOE grant
               DE-FG02-92ER40706}}
\end{center}
\vspace{0.5cm}

\normalsize\sl
\begin{center}
\begin{tabular}{ll}
\kern 0.0em \begin{tabular}{c}
$^{\dagger}$Department of Physics \\
 Bowdoin College  \\
 Brunswick, ME 04011
\end{tabular}    &     \begin{tabular}{c}
                          $^\ddagger$Department of Physics \\
                                 Brandeis University  \\
                               Waltham, MA 02254
                                    \end{tabular}   \\[1.0cm]
\rm naculich@polar.bowdoin.edu &
                   \rm  hriggs@binah.cc.brandeis.edu
\end{tabular} \\[1.0cm]
\end{center}

\rm
\vfill
\begin{center}
{\sc Abstract}
\end{center}

\begin{quotation}
Compact string expressions are found for non-intersecting
Wilson loops in $\sun$ Yang-Mills theory on any surface
(orientable or nonorientable)
as a weighted sum over covers of the surface.
All terms from the coupled chiral
sectors of the $1/N$ expansion of
the Wilson loop expectation values are included.
\end{quotation}
\vfill
{\normalsize \sf November 1994} \hfill

\setcounter{page}{0}
\newpage
\setcounter{page}{1}
\setcounter{section}{0}

The recent interpretation of $\sun$-gauge-theory
Wilson loops on a variety of surfaces as string
amplitudes\cite{gt,minahan,twist}
is a potentially useful prelude to finding a string picture of
two-dimensional hadrons, and, one hopes, to finding a
generalization to four dimensions and real hadrons.
This interpretation divides
the $1/N$ expansion of the exact expressions for
such Wilson loops into two coupled chiral sectors.
If the terms that couple the sectors are neglected, the geometric
interpretation is reasonably straightforward.\cite{twist}
While the structure of $\son$ (and $\spn$) gauge-theory Wilson loops
is similar to a single chiral $\sun$ sector in that it involves
a single sum over Young tableaux, it also includes
features similar to those that couple the $\sun$ chiral sectors.
Nevertheless, a compact and explicit string expression is available for the
entire $1/N$ expansion of non-intersecting $\son$ and $\spn$ Wilson loops,
one that makes their geometric nature transparent.\cite{nrswilson}

In this letter we obtain analogous compact and explicit string expressions
for non-intersecting $\sun$-gauge-theory Wilson loops on arbitrary
(orientable or nonorientable) surfaces,
including all the terms that couple the chiral sectors.
These expressions exhibit the complete set of open string maps
which contribute to the Wilson loop expectation values
in the full coupled theory.

We begin with the analysis of non-intersecting Wilson loops as sums over
partition functions on surfaces with boundary.
Let $Z(\cS; \bU)$ denote the partition function
on an open surface $\cS$
with gauge-field holonomies $U_1,\ldots, U_b$
on the $b$ components of the boundary of $\cS$.
Gauge invariance implies that such partition functions are
expandable on a complete basis of characters of the gauge group
\eq
   \hchi^{\sst SU}_{\dot{R}S}(U) \; ,
\en
with $R$ and $S$ denoting arbitrary tableaux
and $\RS$ denoting the bitableau\cite{king} formed from them
by adjoining the right-justified tableau $R$ with dots in each cell
to the ordinary  left-justified tableau $S$.
As long as the column lengths of $R$ and $S$ are small relative to $N$
there will be no overcounting of representations.
In fact, the only terms that contribute to the $1/N$ expansion are
those in which $R$ and $S$ are finite-cell tableaux\cite{twist}
(finite relative to $N$, which is taken to be large).
The advantage of this basis is that the
expansion coefficients in
\eq
Z(\cS; \bU) = \sum_{R,S} \cZ(\cS; \{\RS,\ldots,\RS\})
\prod_{j=1}^b \hchi_\RS^{\sst SU}(U_j)
\label{expansion}
\en
have exactly calculable values for any surface $\cS$.
The topologically unique surface with $b$ boundary curves,
Euler characteristic $\Eu=2-q-b$, and given orientability
class may be constructed by gluing together $2h+q^{\prime}+b-2$
three-holed spheres with $h\geq 0$ handles and $q^{\prime}\geq 0$ cross-caps,
given that $2h+q^{\prime} = q$. The surface is orientable if
$q^{\prime} =0$, nonorientable otherwise.
Any such construction allows one to evaluate the
partition function,\cite{mr,wit,bt}
\eq
\cZ(\cS; \RS,\ldots, \RS)  = \epsilon^{q^\prime}
(\dim \RS)^\Eu  \exp\left( -\lambda A C_2 (\RS)\over 2N \right),
\label{exactpart}
\en
where $A$ is the area of $\cS$,
$\sqrt{\lambda/N}$ is the gauge coupling constant,
${\rm dim} \RS$ denotes the dimension of the representation $\RS$,
and $C_2(\RS)$ is its quadratic Casimir.
The factor
\eq
\epsilon = \delta_{\overline{\RS},\RS} (-1)^{(r+s)(N-1)}  = \delta_{R,S}
\label{sign}
\en
only appears for nonorientable surfaces.

To make the string interpretation apparent,
we adopt the double symmetric-group-transform basis\cite{twist}
\eq
      \cV_{\kap,\lam}(U) \equiv \sum_{{R\in \cY_r} \atop {S\in \cY_s}}
\hchi_R(\kap) \hchi_S(\lam) \hchi^{{\sst SU}}_{\RS}(U)
\label{double}
\en
in which $\kap$ ($\lam$) denotes
a conjugacy class of $S_r$ ($S_s$),
the symmetric group of permutations of $r$ ($s$) elements,
and $\cY_r$ ($\cY_s$)
denotes the set of Young tableaux with $r$ ($s$) cells.
Due to the orthogonality and completeness of the symmetric group characters,
\eq
\sum_{R\in \cY_r}  \hchi_R (\kap_1) \hchi_R (\kap_2)
          =   C_{\kap_1} \delta_{\kap_1, \kap_2} ,
\qquad
\sum_{\kap \in \cK_r} {1\over C_\kap } \hchi_{R} (\kap) \hchi_{R^\prime} (\kap)
 =   \delta_{R, R^\prime} ,
\label{symorth}
\en
(in which $C_\kap= r!/|\kap|$ for $\kap\in \cK_r$,
the set of conjugacy classes of $S_r$,
with $|\kap|$ denoting the number of permutations in class $\kap$),
the basis (\ref{double}) is also orthogonal and complete
\eq
\int {\rm d} U~\cV_{\kap_1,\lam_1}(U) \cV_{\kap_2,\lam_2}(U^{-1})
= C_{\kap_1\cdot \lam_1} \delta_{\kap_1,\kap_2} \delta_{\lam_1,\lam_2}
          \delta_{r_1,r_2} \delta_{s_1,s_2} \; .
\en
Here $\kap_i \in \cK_{r_i}$ and $\lam_i \in \cK_{s_i}$.
The notation $\kap \cdot \lam$,
for $\kap\in\cK_r$ and $\lam\in\cK_s$,
indicates the conjugacy class in
$S_r \oplus S_s$ composed of the outer
product of elements of $\kap$ and $\lam$.
Note that $C_{\kap\cdot \lam} = C_\kap C_\lam$.

For an arbitrary surface $\cS$\/
the partition function has the expansion
\eq
Z(\cS; \bU )
=  \sum_{r_i \atop s_i }
\sum_{{ {\kap_i \in \cK_{r_i}\atop \lam_i \in \cK_{s_i}}
 \atop {\sct \kern -0.9cm {i=1,\ldots,b}}}}
\cZ(\cS; \{ \kap_1,\lam_1 \}, \ldots, \{ \kap_b,\lam_b \}   )
\prod_{j=1}^{b}          \cV_{\kap_j,\lam_j}(U_j) .
\label{gbexp}
\en
The only nonvanishing contributions occur\cite{wit} when
$\kap_j\in \cK_{r}$ and $\lam_j\in \cK_s$ for all $j$ for
given $r$ and $s$.

For $\cS$ orientable,
Gross and Taylor have argued\cite{twist} that
$\cZ(\cS; \{ \kap_1,\lam_1 \}, \ldots, \{ \kap_b,\lam_b \}   )  $
has a natural interpretation as a weighted sum over $(r+s)$--sheeted
covers of $\cS$,
in which the classes $\kap_1,\ldots,\kap_b \in \cK_r$
describe the boundary covering for the $r$ covering sheets with the same
orientation as $\cS$,
and $\lam_1,\ldots,\lam_b\in\cK_s$ describe the boundary
covering for the $s$ sheets with the opposite orientation as $\cS$.
(The fact that they use a basis of permutations
rather than conjugacy classes only leads to a slight difference as to
whether the sheets that end on a boundary in one orientation sector
are considered to be identical or not\cite{nrswilson}.)

For $\cS$ nonorientable,
one may show that the coefficients in expansion (\ref{gbexp})
have the form of a weighted sum over {\em orientable} covers of $\cS$.
One begins with the observation that
all orientable covers of $\cS$ (with $b$ boundary components)
are covers of $\tcS$, the orientable double cover of $\cS$,
which has $2b$ boundary components.
Thus, each orientable cover of $\cS$
necessarily has an even number $2r$ of sheets,
with boundary coverings specified by the conjugacy classes
$ \{\kap_1,\lam_1\},\ldots,\{\kap_b,\lam_b\}$.
(Here, $\kap_j$ and $\lam_j$ specify the coverings
of two boundary components of $\tcS$ which correspond
to a single boundary component of $\cS$.)
Using this set of covers, a calculation analogous to that in
ref.~\dcite{nrswilson} shows that
$ \cZ(\cS;\{\kap_1,\lam_1\},\ldots,\{\kap_b,\lam_b\}) $
is a weighted sum over orientable covers of $\cS$.
{}From the relation between the expansion coefficients
for any $\cS$,
\eq
  \cZ(\cS;\{\kap_1,\lam_1\},\ldots,\{\kap_b,\lam_b\}) =
\sum_{R,S} \cZ(\cS; \RS,\ldots,\RS)
\prod_{i=1}^b C^{-1}_{\kap_i\cdot\lam_i}\hchi_R(\kap_i) \hchi_S(\lam_i)
\delta_{r_i,r}\delta_{s_i,s} \; ,
\en
one sees that the self-conjugacy of the contributing
representations (\ref{sign}) for $\cS$ nonorientable
leads to the covering constraint $r_i=s_i=r=s$ for all $i$.
Even though $\cS$ is nonorientable,
near the $b$ boundaries a given local orientation
in $\cS$ can be lifted to the sheets of each
connected (orientable, even-sheeted) component
of the cover.
The orientation lifted to the sheets that correspond
to the boundary condition $\lam_j$ will be opposite
that lifted to the sheets corresponding to $\kap_j$.
Therefore, the two-sector structure
of the boundary conditions remains associated with orientability even
when $\cS$ is nonorientable.
However, the two boundary sectors are tied together
in that every connected component of the cover has boundaries corresponding
to both $\lam_j$ and $\kap_j$.
In fact, these results allow one to consistently sew together
covers of nonorientable surfaces with boundary
with covers of orientable surfaces with boundary.

With the string interpretation of the coefficients
$ \cZ(\cS;\{\kap_1,\lam_1\},\ldots,\{\kap_b,\lam_b\}) $
in hand,
we first consider a Wilson loop on a
homologically trivial curve on
an orientable or nonorientable surface $\cS$
({\it i.e.}, the curve divides $\cS$ into two surfaces,
$\cS_a$ and $\cS_b$).
Instead of the standard Wilson loop expectation value
associated with the gauge group representation $\RS$,
\eq
   W_\RS = \int {\rm d}U~ Z(\cS_a; U) \hchi_{\RS}(U) Z(\cS_b; U^{-1}) \; ,
\en
we will take the symmetric-group transforms of such Wilson loops
\eq
    W_{\kap,\lam} = \sum_{{R\in \cY_r \atop S\in \cY_s}}
         \hchi_R(\kap) \hchi_S(\lam) W_{\RS}
\en
as the fundamental objects amenable to a string interpretation.
In order to evaluate
\begin{eqnarray}
 \ds   W_{\kap,\lam} & = & \int {\rm d}U ~Z(\cS_a; U) \cV_{\kap,\lam} (U)
      Z(\cS_b; U^{-1}) \nonumber\\
     & = & \ds \sum_{\mu_a,\nu_a \atop \mu_b, \nu_b}
        \cZ(\cS_a; \{ \mu_a,\nu_a \}) \cZ(\cS_b; \{\mu_b,\nu_b \})
        \int {\rm d} U \cV_{\mu_a,\nu_a}(U)
       \cV_{\kap,\lam}(U)  \cV_{\mu_b,\nu_b} (U^{-1})\nonumber \\
\label{wlfir}
\end{eqnarray}
we need to calculate the product
\eq
   \cV_{\mu,\nu}(U) \cV_{\kap,\lam}(U) =
 \sum_{R_1,S_1 \atop R_2 S_2}
 \hchi_{R_1}(\mu) \hchi_{S_1}(\nu) \hchi_{R_2}(\kap) \hchi_{S_2}(\lam)
 \sum_{R_3,S_3} \multip \hchi^{\sst SU}_{\dR_3S_3}(U) ,
\label{calVprod}
\en
where $\multip$ is the multiplicity of
the representation $\dR_3S_3$ in the $\sun$ tensor product of
$\dR_1S_1$ and $\dR_2S_2$.
If $N$ is sufficiently large,
this multiplicity can be
written as a sum of Littlewood-Richardson coefficients\cite{king}
\eq
   \multip = \sum_{\alpha,\beta} {L_{(R_1/\alpha)(R_2/\beta)}}^{R_3}
                         {L_{(S_1/\beta)(S_2/\alpha)}}^{S_3},
\en
where
\eq
      (R/\alpha) \equiv \sum_D {L_{\alpha D}}^R \; D,
\en
so that
\eq
  \multip = \sum_{{\alpha, \beta \atop D_1,D_2,E_1,E_2}}
    {L_{\alpha D_1}}^{R_1} {L_{\beta D_2}}^{R_2} {L_{D_1 D_2}}^{R_3}
    {L_{\beta E_1}}^{S_1} {L_{\alpha E_2}}^{S_2} {L_{E_1 E_2}}^{S_3} \; .
\en
Using this and the identities\cite{robinson}
\eq
\hchi_R (\kap_1 \cdot \kap_2)
= \sum_{R_1 \in \cY_{r_1} } \sum_{R_2 \in \cY_{r_2} }
{L_{R_1 R_2}}^R \hchi_{R_1} (\kap_1) \hchi_{R_2} (\kap_2)
\en
and
\eq
\sum_{R\in\cY_{r_1+r_2} } \hchi_R(\kap) {L_{R_1 R_2}}^R
= \sum_{\kap_1 \in \cK_{r_1} } \sum_{\kap_2 \in \cK_{r_2} }
   \hchi_{R_1} (\kap_1) \hchi_{R_2} (\kap_2)
                 \delta_{\kap, \kap_1 \cdot \kap_2 }
\en
in expansion~(\ref{calVprod}),
we find that
\eq
\cV_{\mu,\nu}(U) \cV_{\kap,\lam}(U) =
\sum_{{\mu_1,\mu_2} \atop {{\nu_1,\nu_2}\atop {\kap_2,\lam_2}}}
 C_{\mu_1\cdot \nu_1}
\delta_{\mu, \mu_1\cdot \mu_2}  \delta_{\nu, \nu_1\cdot \nu_2}
\delta_{\kap, \nu_1\cdot \kap_2}  \delta_{\lam, \mu_1\cdot \lam_2}
\cV_{\mu_2\cdot \kap_2, \nu_2 \cdot \lam_2}(U)
\en
After further simplication, equation~(\ref{wlfir}) becomes
\eq
   W_{\kap,\lam} = C_{\kap\cdot \lam}
\sum_{ { {\lam_e,\lam_o}\atop {\kap_e,\kap_o} }}
\delta_{\kap, \kap_e\cdot \kap_o} \delta_{\lam, \lam_e\cdot \lam_o}
\sum_{{p_e}\atop {p_o}}
\sum_{{\pi_e\in \cK_{p_e}}\atop{\pi_o\in\cK_{p_o}}}   C_{\pi_e\cdot\pi_o}
      \cZ(\cS_a; \{ \lam_e \cdot \pi_e , \kap_o\cdot \pi_o \})
      \cZ(\cS_b;\{ \kap_e\cdot \pi_e, \lam_o\cdot \pi_o\} )
\label{expression}
\en
With the coefficients $\cZ(\cS)$
interpreted as weighted sums over (orientable) covers of the
orientable or nonorientable surfaces $\cS_a$ and $\cS_b$,
this formula provides a compact and geometrically transparent
expression for the Wilson loop as a sum over maps from surfaces
$\cW$ with boundary to $\cS$,
with the boundary of each $\cW$ mapped
to the curve in $\cS$ on which the Wilson loop is defined.
For each pair of divisions $\kap=\kap_e\cdot \kap_o$ and
$\lam=\lam_e\cdot\lam_o$ into subcycle conjugacy
classes, we sew the orientation-preserving
covers of $\cS_a$ (with $\lam_e\cdot\pi_e$
describing the boundary covering) to the orientation-preserving covers
of $\cS_b$ (with boundary covering $\kap_e\cdot\pi_e$)
by letting the $r_e$ sheets of $\lam_e$
end on the Wilson loop on one side,
by letting the $s_e$ sheets of $\kap_e$
end on the Wilson loop on the other side,
and by sewing the $p_e$ sheets corresponding to $\pi_e$ on
either side together in all possible ways, as described in section
{\it 3} of ref.~\dcite{nrswilson}.
Similarly we sew the orientation-reversing
covers of $\cS_a$
(with boundary covering $\kap_o\cdot\pi_o$)
to the orientation-reversing covers
of $\cS_b$ (with boundary covering $\lam_o\cdot\pi_o$)
by letting the $r_o$ sheets of $\lam_o$
end on the Wilson loop on one side,
by letting the $s_o$ sheets of $\kap_o$
end on the Wilson loop on the other side,
and by sewing the $p_o$ sheets corresponding to $\pi_o$ on
either side together in all possible ways,
as was done in the other sector.
The presence of the factor
$C_{\kap\cdot\lam}$ simply means that we should consider sheets that
end on the Wilson loop as distinct rather than identical when computing
the proper weight to attach to each surface.

{}From equation~(\ref{expression}),
one easily deduces the relation
$r_a - s_a = r_b - s_b + r_\lambda - r_\kappa$
between the number of orientation-preserving ($r_a$)
and orientation-reversing ($s_a$) sheets over $\cS_a$,
and the number of orientation-preserving ($r_b$)
and orientation-reversing ($s_b$) sheets over $\cS_b$,
in terms of $r_\lambda$ and $r_\kappa$,
where $ \kappa \in S_{r_\kappa}$ and $ \lambda \in S_{r_\lambda}$.

The special case $W_{\kap, 0}$
\eq
  W_{\kap,0} = C_\kap  \sum_{\kap_e,\kap_o} \delta_{\kap, \kap_e \cdot \kap_o}
\sum_{{p_e=0}\atop{p_o=0}}
 \sum_{{\pi_e\in\cK_{p_e}}\atop{\pi_o\in\cK_{p_o}}} C_{\pi_e\cdot\pi_o}
   \cZ(\cS_a; \pi_e, \kap_o\cdot \pi_o) \cZ(\cS_b; \kap_e\cdot \pi_e, \pi_o)
\en
illustrates the structural parallel with the result\cite{nrswilson}
\eq
  W_\kap =
C_\kap \sum_{\kap_1,\kap_2} \delta_{\kap,\kap_1 \cdot \kap_2}
           \sum_{p} \sum_{\pi\in \cK_{p}} C_{\pi}
\cZ(\cS_a; \kap_1\cdot \pi) \cZ(\cS_b; \kap_2 \cdot \pi) \; .
\en
for the gauge groups $\son$ and $\spn$.

For a homologically {\em nontrivial} curve
which cuts the closed surface $\cS_c$ into an open surface $\cS$
with two boundaries,
the $\sun$ Wilson loop expectation value is
\eq
 \begin{array}{l}
\ds  W_{\kap,\lam} =
\int {\rm d}U~ Z(\cS; U,U^{-1}) \cV_{\kap,\lam}(U)  \\[1.0cm]
\ds  = C_{\kap\cdot \lam} \sum_{{\kap_e,\kap_o}\atop {\lam_e,\lam_o}}
 \delta_{\kap,\kap_e \cdot \kap_o} \delta_{\lam,\lam_e\cdot\lam_o}
   \sum_{p_e \atop p_o} \sum_{\pi_e\in \cK_{p_e}\atop \pi_o\in \cK_{p_o}}
C_{\pi_e \cdot \pi_o}
 \cZ(\cS; \{ \lam_e \cdot \pi_e, \kap_o \cdot \pi_o\},
           \{ \kap_e \cdot \pi_e, \lam_o \cdot \pi_o\}) .
\end{array}
\en
Again, this provides a compact prescription for sewing covers of
$\cS$ together along the two boundaries to form maps from string worldsheets
$\cW$ to $\cS_c$ with the boundary of $\cW$ mapped to the Wilson loop.
Note that this Wilson loop  vanishes
unless $\kap$ and $\lam$ are conjugacy classes of the same
symmetric group
and have subdivisions into cycles of the same length,
since $\lam_e$ ($\lam_o$)
must have the same length as $\kap_e$ ($\kap_o$).

For a nonorientable curve\cite{nrswilson} on a nonorientable surface
$\cS_n$,
the Wilson loop expectation value is given by
\eq
W_{\kap,\lam}
=  \int {\rm d}U~ Z(\cS; U^2) \cV_{\kap,\lam}(U)
=  \sum_{p \atop r} \sum_{\pi\in \cK_{p}\atop \rho\in \cK_{r}}
 \cZ(\cS; \{ \pi, \rho\})   M_{ \{\kap, \lam\}, \{\pi, \rho\} }
\en
where
\eq
M_{\{\kap, \lam\},\{\pi, \rho\} }
= \sum_{P,Q,R,S} \hchi_P(\pi) \hchi_Q (\rho) \hchi_R(\kap) \hchi_S(\lam)
\left(   {N_{\dP Q, \dP Q}^+}^{\dR S}
       - {N_{\dP Q, \dP Q}^-}^{\dR S} \right)
\en
is an integer that plays the same combinatorial
role for sewing together covers
of a single boundary (that is to be glued to itself to
form $\cS_n$ from $\cS$) as $C_{\pi_e \cdot \pi_o}$ does for sewing
together the covers over two glued-together boundaries.

The problem of finding analogous compact formulae for intersecting
Wilson loops that include the coupling terms in a way that makes the
geometry transparent remains a challenge.

\noindent{\bf Acknowledgment:}
It is a pleasure to thank S. T. Fisk and J. A. Wood for
some very useful comments.

\end{document}